\documentclass[12pt]{article}
\usepackage{array,natbib,url,graphicx,amsmath,amssymb,latexsym}
\newtheorem{theorem}{Theorem}
\newtheorem{lemma}{Lemma}
\newtheorem{remark}{Remark}
\newtheorem{corollary}{Corollary}

\begin{document}
\renewcommand{\baselinestretch}{1.6}
%\lhead[\fancyplain{} \leftmark]{}
%\chead[]{}
%\rhead[]{\fancyplain{}\rightmark}
%\cfoot{}
%\headrulewidth=0pt
\markright{
%\hbox{\footnotesize\rm Statistica Sinica
%{\footnotesize\bf ??}(200?), 000-000}\hfill
}
\markboth{\hfill{\footnotesize\rm HENG LIAN
}\hfill}
{\hfill {\footnotesize\rm VARIANCE FUNCTION ESTIMATION} \hfill}
\renewcommand{\thefootnote}{}
$\ $\par
\fontsize{10.95}{14pt plus.8pt minus .6pt}\selectfont
\vspace{0.8pc}
\centerline{\large\bf Nonparametric Estimation of Variance Function }
\vspace{2pt}
\centerline{\large\bf for Functional Data}
\vspace{.4cm}
\centerline{Heng Lian}
\vspace{.4cm}
\centerline{\it Division of Mathematical Sciences,}
\centerline{\it School of Physical and Mathematical Sciences,}
\centerline{\it Nanyang Technological University, Singapore 637371.}
\vspace{.55cm}
\fontsize{9}{11.5pt plus.8pt minus .6pt}\selectfont

\begin{quotation}
\noindent {\it Abstract:}
This article investigates nonparametric estimation of variance functions for functional data when the mean function is unknown. We obtain asymptotic results for the kernel estimator based on squared residuals. Similar to the finite dimensional case, our asymptotic result shows the smoothness of the unknown mean function has an effect on the rate of convergence. Our simulaton studies demonstrate that estimator based on residuals performs much better than that based on conditional second moment of the responses. \par

\vspace{9pt}
\par
\end{quotation}\par

\fontsize{10.95}{14pt plus.8pt minus .6pt}\selectfont
\section{Introduction}
Recently, there has been increased interest in the statistical modelling of functional data. In many experiments, functional data appear as the basic unit of observations. As a natural
extension of the multivariate data analysis, functional data analysis provides valuable
insights into these problems. Compared with the discrete multivariate analysis, functional
analysis takes into account the smoothness of the high dimensional covariates, and often
suggests new approaches to the problems that have not been discovered before. Even for
nonfunctional data, the functional approach can often offer new perspectives on the old
problem.

The literature contains an impressive range of functional analysis tools for various problems
including exploratory functional principal component analysis, canonical correlation
analysis, classification and regression. Two major approaches exist. The more traditional
approach, carefully documented in the monograph \cite{ramsey05}, typically
starts by representing functional data by an expansion with respect to a certain basis, and
subsequent inferences are carried out on the coefficients. The most commonly utilized basis
include B-spline basis for nonperiodic data and Fourier basis for periodic data. Another line
of work by the French school \cite{ferraty02}, taking a nonparametric point of view, extends the traditional nonparametric techniques, most notably the kernel estimate, to the functional case. Some
theoretical results are also obtained as a generalization of the convergence properties of the
classical kernel estimate.

The functional nonparametric regression model, introduced in \cite{ferraty02}, is defined as
\begin{equation}\label{model}
Y_i=m(X_i)+\sqrt{v(X_i)}\epsilon_i,
\end{equation}
where we emphasized the heterogeneity of the regression model which is the focus of this article. We assume that $\epsilon_i$'s are random variables with $E(\epsilon_i|X_i)=0$ and $Var(\epsilon_i|X_i)=v(X_i)$. The covariates $X_i$ are assumed to belong to some semi-metric vectorial space $\mathcal{H}$ determined by the semi-metric $d(.,.)$. Unlike many previous nonparametric functional regression studies \cite{ferraty04,masry05,lian07} which focused on estimating the mean function $m$, here we are interested in estimating $v$ when $m$ is unknown, and thus the mean function only plays the role of a nuisance parameter. 

Variance function estimation has received much attention since the 1980's when it was required for confidence interval construction for the mean function, and \cite{muller87} discussed some utility of it in obtaining more efficient estimators of the mean function. There are two main approaches to variance function estimation.  In \cite{muller87,muller93}, the variance function was estimated directly from local contrasts of the responses. More recently, \cite{brown07,wang08} has obtained minimax convergence rates based on local difference and \cite{cai09} further extended this to multivariate regression. These asymptotic theory were developed based on fixed covariates on a grid and it is not straightforward to extend to the case with random covariates. For our functional data analysis, it is not clear how to define a grid on the semi-metric space $\mathcal{H}$. A different direction was taken in \cite{hall89}, where the variance function was estimated by a weighted smoothing of squared residuals after a fit for the mean function was obtained. This approach was also considered in \cite{fan98} using local polynomial regression. Finally, we mention the adaptive estimation of variance function in \cite{cai08} by thresholding of wavelet coefficients.

In the following sections, we adapt the idea of variance estimation in nonparametric regression based on squared residuals to the functional setting. In Section 2, we review the functional nonparametric regression model in a semi-metric functional vectorial space. Then we introduce functional nonparametric variance estimation in this general setting and describe the asymptotic results for our kernel-type estimator. We also discuss the effect of unknown mean function on the variance estimator and relate it to the finite-dimensional case. In Section 3, we carry out a simulation study to demonstrate that the residual-based estimator is more efficient than the estimator based on nonparametric regression on the squared responses. Finally, we illustrate the approach on the popular spectrometric data for predicting the fat content. The technical proofs for our asymptotic results are deferred to the appendix.   

\section{Nonparametric Functional Variance Estimation}

In the functional nonparametric regression model (\ref{model}) presented originally in \cite{ferraty02}, the mean function is estimated by a kernel-type estimator
\[\hat{m}(x)=\frac{\sum_{i=1}^nK(d_m(x,X_i)/h_m)Y_i}{\sum_{i=1}^nK(d_m(x,X_i)/h_m)},
\]
where $Y_i$ is the real-valued responses and $h_m$ is the bandwidth used for estimating the mean function. Note that we use $d_m$ to denote the semi-metric for mean function estimation as we will use a different semi-metric for variance function estimation.
Denote $R(X,Y)=(Y-m(X))^2$. Since under model (\ref{model}), we have $E(R(X,Y)|X)=v(X)$, a natural kernel-type estimator for $v(x)$ (when the mean function is known) is 
\begin{equation}\label{var.est}
\hat{v}(x)=\frac{\sum_{i=1}^nK(d_v(x,X_i)/h_v)R_i}{\sum_{i=1}^nK(d_v(x,X_i)/h_v)},
\end{equation}
where $R_i=(Y_i-m(X_i))^2$ and $h_v$ is the chosen bandwidth of the kernel. Note that the semi-metric $d_v$ used for estimating the variance function is in general different from the semi-metric $d_m$ used in estimating the mean function. Using different semi-metrics is important in some cases as demonstrated in our experiment with spectrometric data later. Although we could use different kernels for the mean and variance functions, we choose to use the same kernel here mainly for notational simplicity. 

In practice, the mean function $m(\cdot)$ is typically unknown and a natural approach is to replace $m$ by the nonparametric estimator $\hat{m}$. Equivalently, we replace $R_i$ by $\hat{R}_i=(Y_i-\hat{m}(X_i))^2$ in (\ref{var.est}).

Although only independent data are considered in our simulations and real data application, for our asymptotic analysis, we will present our results in a more general context by considering a strongly mixing sequence $\{(X_i, Y_i),i=1,\ldots,n\}$. Our asymptotic result is stated for a fixed $x\in \mathcal{H}$. %for convergence in probability. %Uniform convergence requires additional topological assumption on the semi-metric space and the reader can refer to \cite{ferratyerror} for the assumptions.

Following the notations in \cite{ferraty06}, we have
\begin{eqnarray*}
\Delta^m_i&=&\frac{K(d_m(x,X_i)/h_m)}{EK(d_m(x,X_i)/h)_m}\\
r^m_1&=&\sum_{i=1}^n\Delta^m_i/n\\
r^m_2&=&\sum_{i=1}^n Y_i\Delta^m_i/n\\
\Delta^v_i&=&\frac{K(d_v(x,X_i)/h_v)}{EK(d_v(x,X_i)/h_v)}\\
r^v_1&=&\sum_{i=1}^n\Delta^v_i/n\\
r^v_2&=&\sum_{i=1}^n (Y_i-\hat{m}(X_i))^2\Delta^v_i/n\\
\end{eqnarray*} 
so that $\hat{m}(x)=r^m_2/r^m_1$ and $\hat{v}(x)=r^v_2/r^v_1$. For notational simplicity, in the rest of the article, we denote $m_i=m(X_i), \hat{m}_i=\hat{m}(X_i), v_i=v(X_i), \hat{v}_i=\hat{v}(X_i).$ We also set $w_{ij}=K(d_m(X_i,X_j)/h_m)/\sum_kK(d_m(X_i,X_k)/h_m)$ so that $\hat{m}_i=\sum_jw_{ij}Y_j$.

Similar to \cite{ferraty04,ferraty06}, the rate of convergence of $\hat{v}(x)$ will critically depend on the quantities $s_n^m$ and $s_n^v$ defined by
\begin{eqnarray}
s_n^m&=&\max\{s^m_{n,1},s^m_{n,2},s^m_{n,3},s^m_{n,4}\}\nonumber\\
s_n^v&=&\max\{s^v_{n,1},s^v_{n,2},s^v_{n,3},s^v_{n,4}\}\nonumber\\
s^m_{n,1}&=&\sum_{i=1}^n\sum_{j=1}^n|Cov(\Delta^m_i,\Delta^m_j)|\label{smn1}\\
s^m_{n,2}&=&\sum_{i=1}^n\sum_{j=1}^nE|Cov(\Delta^m_i\epsilon_i,\Delta^m_j\epsilon_j|X_1^n)|\label{smn2}\\
s^m_{n,3}&=&\sum_{i=1}^n\sum_{j=1}^n|Cov(\Delta^m_im_i,\Delta^m_jm_j)|\label{smn3}\\
s^m_{n,4}&=&n|E\sum_{i=1}^n\sum_{j=1}^n\Delta^v_iw_{ij}\epsilon_i\epsilon_j|\label{smn4}\\
s^v_{n,1}&=&\sum_{i=1}^n\sum_{j=1}^n|Cov(\Delta^v_i,\Delta^v_j)|\label{svn1}\\
s^v_{n,2}&=&\sum_{i=1}^n\sum_{j=1}^nE|Cov(\Delta^v_i\epsilon_i,\Delta^v_j\epsilon_j|X_1^n)|\label{svn2}\\
s^v_{n,3}&=&\sum_{i=1}^n\sum_{j=1}^n|Cov(\Delta^v_iv_i,\Delta^v_jv_j)|\label{svn3}\\
s^v_{n,4}&=&\sum_{i,j,k,l=1}^nE|Cov(\Delta_i^vw_{ij}\epsilon_i\epsilon_j,\Delta_k^vw_{kl}\epsilon_k\epsilon_l|X_1^n)|\label{svn4}
\end{eqnarray}
where in some of the expressions above, the covariances are conditioned on observed covariates $X_1^n=\{X_1,\ldots,X_n\}$.
 
We follow \cite{ferraty06} and impose the following condition on the kernel function
\begin{equation}\label{assum:kernel}
\mbox{$K$ is supported on $[0,1]$, bounded and bounded away from zero on $[0,1]$}.
\end{equation}

As the case for mean function estimation, we need the following regularity conditions
\begin{equation}\label{assum:lip}
|m(x_1)-m(x_2)|\le Cd_m(x_1,x_2)^\alpha, |v(x_1)-v(x_2)|\le Cd_v(x_1,x_2)^\beta, \alpha>0, \beta>0.
\end{equation}
In \cite{ferraty04,ferraty06}, moment conditions are directly assumed on the response $Y$. We figure that it is more natural to impose the moment condition on the error
\begin{equation}\label{assum:error}
\exists p\ge 4, E|\epsilon|^p<\infty .
\end{equation}

For uniform convergence over a compact neighborhood $\mathcal{C}$ of $\mathcal{H}$ containing $x$ for the mean function, which is needed in the proof below, we assume that $\mathcal{C}$ can be written as, for any $l>0$,
\begin{equation}\label{assum:top}
\mathcal{C}=\sum_{k=1}^\tau\mathcal{B}(t_k,l), \mbox{ with } \tau l^a=C \mbox{ for some } a>0, C>0.
\end{equation}
This condition is exactly the same as that in \cite{ferratyerror}, and interested readers can find some related discussions there.

%Next, we assume the data come from some stationary strong mixing processes
%\[\alpha(n)\le cn^{-a}, \exists \theta>0, s_n^{-(a+1)}=o(n^{-\theta})\]
%This condition is also assumed in Chapter 11 of \cite{ferraty06} where geometric mixing is also studied. We omit this and the reader can find much information there. 

Now we are ready to state our main result:
\begin{theorem}\label{thm:1}
Under the conditions (\ref{assum:kernel})-(\ref{assum:top}), for a fixed $x\in \mathcal{H}$, we have
\[|\hat{v}(x)-v(x)|=O\left(h_m^{2\alpha}+\frac{{s^m_n}\log n}{n^2}+h_v^\beta+\frac{\sqrt{s^v_n\log n}}{n}\right)\;\; in\;\; probability.\]
\end{theorem}
\begin{remark}
In \cite{ferraty04,ferraty06}, the asymptotic results are stated as almost complete convergence, which is stronger than convergence in probability. The difficulty of proving stronger convergence for our variance estimator comes from the appearance of U-type-statistics in the expressions in the proof, thus we settle with weaker type of convergence here.
\end{remark}
\begin{remark}
In \cite{ferraty06}, it was discussed in details how $s_n^m$ depends on the following two quantities: $\phi_m(h):=P(d_m(x,X)\le h)$ and $\psi_m(h)=P(d_m(x-X_1)\le h, d_m(x,X_2)\le h)$ for strongly mixing data sequences. Those results can be adapted for our purposes. For example, for the independent and identically distributed data, as shown in the appendix, we have $s_n^m=O(n/\phi_m(h_m))$ and $s_n^v=O(n/\phi_v(h_v))$ with $\phi_v(h)=P(d_v(x,X)\le h)$. Thus in the i.i.d. case we have the following direct consequence.
\end{remark}
\begin{corollary}\label{cor:1}
Under the conditions (\ref{assum:kernel})-(\ref{assum:top}), assuming in addition the data $\{(X_i,Y_i),i=1,\ldots,n\}$ are i.i.d. and the bandwidths are chosen such that $h_m\rightarrow 0, h_v\rightarrow 0, \; n\phi_m(h_m)\rightarrow\infty, n\phi_v(h_v)\rightarrow\infty$, we have
\[|\hat{v}(x)-v(x)|=O\left(h_m^{2\alpha}+\frac{\log n}{n\phi_m(h_m)}+h_v^\beta+\sqrt{\frac{\log n}{n\phi_v(h_v)}}\right)\;\; in\;\; probability.\]
\end{corollary}
\begin{remark}
From the corollary, we can observe some interesting effect of unknown mean for variance  function estimation. For simplicity and specificity, assume that $X$ is of fractal order $d$ with respect to both $d_m$ and $d_v$, i.e. $\phi_m(h)\sim\phi_v(h)\sim h^d$. It was shown in \cite{ferraty06} Lemma 13.6 that if $\mathcal{H}$ is a separable Hilbert space with semi-metric defined by the projection onto the first $d$ elements of an orthonormal basis, then $\phi(h)\sim h^d$. This is also true for $d$-dimensional regression (i.e., $\mathcal{H}=R^d$). With $h_m\sim(\log n/n)^{1/(2\alpha+d)}, h_v\sim(\log n/n)^{1/(2\beta+d)}$, we obtain the rate of convergence $\max\{(\log n/n)^{2\alpha/(2\alpha+d)}, (\log n/n)^{\beta/(2\beta+d)}\}$. If $2\alpha/(2\alpha+d)\ge\beta/(2\beta+d)$, the rate becomes $(\log n/n)^{\beta/(2\beta+d)}$. This rate is the same as the rate obtained when the mean function $m(.)$ is known. Thus we observe that when the mean function is smooth enough, it has no effect on variance function estimation, while its effect cannot be ignored for less smooth mean functions. In particular, it can be easily verified that $2\alpha/(2\alpha+d)\ge\beta/(2\beta+d)$ is true as soon as $\alpha\ge d/2$. This results is the same as what was observed in \cite{hall89} for one-dimensional regression where the author observed that the mean has no effect on variance function estimation as long as $\alpha\ge 1/2$ (the last sentence in section 2.2 of \cite{hall89}).
\end{remark}

\begin{remark}
The simple relationship $v(x)=E(Y^2|X=x)-(E(Y|X=x))^2$ motivates the direct estimator based on estimating conditional expectation of squared responses and setting $\hat{v}(x)=\hat{s}(x)-\hat{m}^2(x)$ where $\hat{s}(x)$ is the nonparametric kernel-type estimate of $E(Y^2|X=x)$. This estimator is briefly mentioned in \cite{ferraty07}. It can be shown that this estimator has the same convergence rate as above. However, in one-dimensional case, \cite{fan98} pointed out the direct method can create a very large bias. The intuitive explanation provided for the large bias is that the direct estimator is obtained when replacing $\hat{R}_i=(Y_i-\hat{m}(X_i))^2$ in the residual-based method by $(Y_i-\hat{m}(x))^2$. This explanation also applies to our functional context. In our simulation study to be presented next, it is clear that the performance of the direct method is much worse than the residual based method.
\end{remark}

\section{Experiments}
\subsection{Simulation Study}
We now consider in this section the finite sample performance of our variance estimator and also compare the results with the direct squared responses based method. We use three examples with different mean and variance functions to illustrate their performances. For each example, 100 simulations are performed with $n=200$ data points generated in each simulation. In all three examples, $X_i$ is a random function supported on $[-1,1]$. 

For the first example, we set 
\[m(x)=0, v(x)=\int_{-1}^1|\cos x(t)|\,dt,\]
and the $X_i$'s are generated as realizations of Brownian Motion starting at time $t=-1$ with random start point $x(-1)$ distributed as uniform random variables on $[-1,1]$. For the second example, we have
\[m(x)=\int_{-1}^1tx(t)\,dt, v(x)=\int_{-1}^1|t|x^2(t)\,dt,\]
and the $X_i$'s are generated the same way as in the first example.
For the third example, we follow \cite{ferraty07} and set
\[m(x)=\int_{-1}^1|x'(t)|(1-\cos(\pi t))dt, v(x)=\int_{-1}^1|x'(t)|(1+\cos(\pi t))dt.\]
The random curves in this example are simulated from
\[X(t)=\sin(\omega t)+(a+2\pi)t+b, \omega\sim Unif(0,2\pi), a,b\sim Unif(0,1).\]
The simulations are performed in R with the publicly available \textbf{npfda} package (\url{http://www.lsp.ups-tlse.fr/staph/npfda/}). The default quadratic kernel is used in the implementation. The bandwidths $h_m$ and $h_v$ are chosen using cross-validation. The choice of semi-metric is in general a difficult problem. In our current simulations, their choices are suggested by our knowledge of the true mean and variance functions. Thus for the first two examples, we use $d_m(x_1,x_2)=d_v(x_1,x_2)=\int_{-1}^1(x_1(t)-x_2(t))^2dt$ and we use $d_m(x_1,x_2)=d_v(x_1,x_2)=\int_{-1}^1(x_1'(t)-x_2'(t))^2dt$ for the third example. Our simulation also shows that these choices of semi-metrics are the best among semi-metrics based on different orders of derivatives (results not represented here). For evaluation of performance, we adopt the discrete mean squared error
\[MSE=\frac{1}{n}\sum_{i=1}^n(\hat{v}(X_i)-v(X_i))^2.\]
We report in Table 1 the median MSE for variance function estimators based on 100 simulations. It is easily seen from the table that and the  residual based two-step method performs much better than the direct method in terms of MSE, except in the first example with constant mean function, which is as expected.

\begin{table}
  \caption{Simulation results (MSE) for comparing two variance function estimators.}
{\begin{tabular}{@{}lcccccc}\hline
   Estimators
  & Example 1 & Example 2  & Example 3         \\
\hline
   residual based method &0.10 & 0.27 & 4.37  \\
   direct method  & 0.10 & 0.38 & 19.24  \\
   \hline
  \end{tabular}}
      \label{symbols}
\end{table}

\subsection{Illustration with Chemometric Data}
We illustrate our approach on the real chemometric dataset, which contains 215 spectra of light absorbance for meat samples as functions of the wavelengths. Because of the denseness of wavelengths at which the measurements are made, the subjects are naturally treated as continuous curves. This dataset has been previously used in nonparametric regression studies where the covariate is the spetra curve and the response is the percentage of fat content in the piece of meat \cite{ferraty02,ferraty06,ferraty07}. We will estimate the variance function for this regression problem. Previous study suggested that for mean function estimation, taking as the semi-metric the $L_2$ distance between the second derivatives of the spetra gives favorable result, thus this semi-metric is used for mean function estimation. As in previous studies, we train on the first 150 spectra and use the rest as validation. We examine the estimation accuracy of the variance function for semi-metrics defined as $L_2$ distance between the curves using different orders of derivatives, measured as mean squared error 
\[ MSE=\frac{1}{65}\sum_{i=151}^{215}(\hat{R}_i^2-\hat{v}(X_i))\]
and find that using $L_2$ distance between 1st derivatives gives the best result. The estimated variance function value and squared residuals for the validation data are shown in Fig. \ref{fig:chemo}, giving a MSE of 33.18. Heterogeneity of the problem are clearly seen from the figure.

\begin{figure}
\begin{center}

\resizebox*{8cm}{!}{\includegraphics{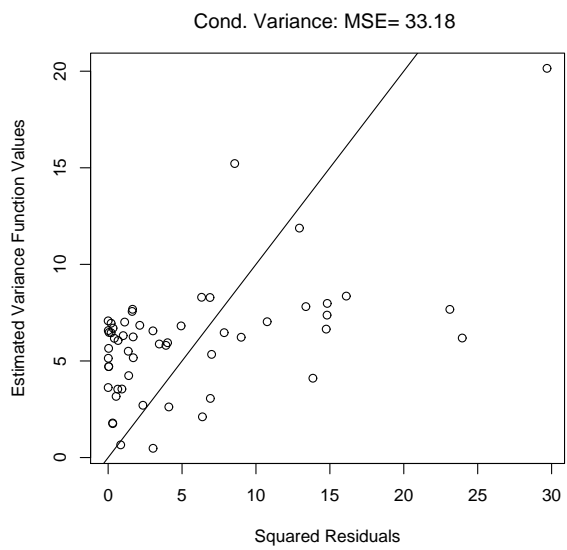}}%
\caption{\label{fig:chemo} Estimated variance function vs. squared residuals on validation data.}%
\label{fig:chemo}
\end{center}
\end{figure}

\section{Conclusion}
In this article, we study the problem of nonparametrically estimating variance function in functional data analysis. We derived the asymptotic property for the squared residuals based estimator and its superiority to the direct squared responses based method is demonstrated through simulations. Our asymptotic result shows an interesting interaction between the smoothness of the mean function and that of the variance function. Finally, we show there exists clear heterogeneity in the regression problem for the chemometric data as an illustration.

\section*{Acknowledgements}
This research is funded by Singapore MOE Tier 1.

\section*{Appendix}
First, we make the remark that under condition (\ref{assum:lip}), we can assume $m(x)\le M_m$ and $v(x)\le M_v$, that is the mean and variance functions are bounded, without loss of generality. The reason is that we always consider only values of both functions inside a compact neighborhood of the fixed $x$. As an illustration, in the definition of the estimator $\hat{v}(x)$, $\Delta_i^v>0$ only when $d_v(x,X_i)\le h_v$, so the sum over $i$ is only for all $X_i$'s contained in a neighborhood of $x$.

To make the presentation clear, we first state the asymptotics for mean function estimation in a Lemma. Note all asymptotic orders obtained below are in the sense of convergence in probability.
\begin{lemma}\label{lem:mean}
Under conditions (\ref{assum:kernel})-(\ref{assum:error}), we have 
\begin{eqnarray}
|r_1^m(x)-1|&=&O(\sqrt{s_n^m\log n/n^2})\label{mean1}\\
|r_2^m(x)-m(x)|&=&O(h^\alpha_m+\sqrt{s_n^m\log n/n^2})\label{mean2}\\
|E(r_2^m/r_1^m)-m(x)|&=&O(h^\alpha_m+\sqrt{s_n^m\log n/n^2})\label{mean3}\\
|r_2^m/r_1^m-E(r_2^m/r_1^m)|&=&O(h^\alpha_m+\sqrt{s_n^m\log n/n^2}).\label{mean4}
\end{eqnarray}
If in addition, condition (\ref{assum:top}) is satisfied, the above convergence is uniform over a compact neighborhood of $x$ in $\mathcal{H}$. 
\end{lemma}
\textbf{Proof:}
The proofs of (\ref{mean1}) and (\ref{mean2}) are similar to that contained in \cite{ferraty04,ferraty06}. On one hand, the proof is simplified by the observation that we only require convergence in probability. On the other hand, the fact that we impose  conditions directly on the errors instead of the responses make the proof slightly more complicated. Equations (\ref{mean3}) and (\ref{mean4}) are direct consequences of the first two equations and the last statement of the lemma follows from \cite{ferratyerror}. We only show (\ref{mean2}) below.

The bias $|Er_2^m(x)-m(x)|=O(h_m^\alpha)$ is shown exactly as in \cite{ferraty04}. For variance calculation, we have
\begin{eqnarray*}
Var(r_2^m-E(r_2^m|X_1^n))&=&E[Var(r_2^m-E(r_2^m|X_1^n)|X_1^n)]\\
&=&E[Var(\frac{1}{n}\sum_i\Delta_i^mv_i\epsilon_i|X_1^n)]\\
&\le&\frac{M_v^2}{n^2} s_{n,2}^m=O(\frac{1}{n^2} s_{n,2}^m).
\end{eqnarray*}
Similarly, $Var( E(r_2^m|X_1^n)-Er_2^m )=O(s_{n,3}^m/n^2)$ using equation (\ref{smn3}). Since $Var(r_2^m)=Var(r_2^m-E(r_2^m|X_1^n))+Var( E(r_2^m|X_1^n)-Er_2^m )=O(s_n^m/n^2), (\ref{mean2})$ follows from the Markov inequality.
$\Box$

\textbf{Proof of Theorem \ref{thm:1}: }

Using the decomposition
\[\hat{R}_i=(Y_i-\hat{m}_i)^2=v_i+2\sqrt{v_i}(m_i-\hat{m}_i)\epsilon_i+(m_i-\hat{m}_i)^2+v_i(\epsilon_i^2-1)=:A_i+B_i+C_i+D_i.\]
and similar to the proof of Lemma \ref{lem:mean}, we have
\begin{equation}\label{combine1}
|r_1^v-1|=O(\sqrt{s^v_{n,1}\log n/n^2}),
\end{equation} and we only need to show that 
\[
r_2^v=\sum_i\Delta_i^v(A_i+B_i+C_i+D_i)/n=O(h_v^\beta+\sqrt{s^v_{n}\log n/n^2}+h_m^{2\alpha}+s_n^m\log n/n^2).
\]
Using conditions (\ref{assum:kernel})-(\ref{assum:error}), we have 
\begin{equation}\label{combine2}
\sum_i\Delta_i^v(A_i+D_i)/n-v(x)=O(h_v^\beta+\sqrt{(s^v_{n,2}+s^v_{n,3})\log n/n^2}),
\end{equation}
following the same steps as the proof of (\ref{mean2}). Also,
\begin{eqnarray}\label{combine3}
\sum_i\Delta_i^vC_i/n&=&\sum_i(m_i-\hat{m}_i)^2\Delta_i^v/n\nonumber\\
&\le&\sup_i(m_i-\hat{m}_i)^2 r_1^v=O(h_m^{2\alpha}+s_n^m\log n/n^2),
\end{eqnarray}
where the supremum over $i$ obeys the same rate as for a fixed $x$  because we can take only $i$ such that $d_v(x,X_i)\le h_v$, which is contained in any fixed compact neighborhood of $x$ and note the final statement of Lemma \ref{lem:mean}.

Finally, the term $\sum_i\Delta_i^vB_i/n$ is dealt with in Lemma \ref{lem:B}. The theorem is proved combining the following lemma with (\ref{combine1}), (\ref{combine2}) and (\ref{combine3}).

\begin{lemma}\label{lem:B}
In the context of Theorem \ref{thm:1}, we have $\sum_i\Delta_i^vB_i/n=O(\sqrt{(s^v_{n,2}+s^v_{n,4})\log n/n^2}+s^m_{n,4}/n^2)$.
\end{lemma}
\textbf{Proof:}
Writing 
\begin{eqnarray}\label{Bdecomp}
\sum_i\Delta_i^vB_i/n&=&\frac{2}{n}\sum_i\Delta_i^v\sqrt{v_i}(m_i-E\hat{m}_i)\epsilon_i+\frac{2}{n}\sum_i\Delta_i^v\sqrt{v_i}(E\hat{m}_i-\hat{m}_i)\epsilon_i\nonumber\\
&=&\frac{2}{n}\sum_i\Delta_i^v\sqrt{v_i}(m_i-E\hat{m}_i)\epsilon_i+\frac{2}{n}\sum_i\Delta_i^v\sqrt{v_i}\epsilon_i(E(\sum_jw_{ij}m_j)-\sum_j(w_{ij}m_j))\nonumber\\
&&-\frac{2}{n}\sum_i\Delta_i^v\sqrt{v_i}\epsilon_i(\sum_j(w_{ij}\sqrt{v_j}\epsilon_j))
=:F+G+H.
\end{eqnarray}
We have $E(F)=E(F|X_1^n)=0$ and 
\begin{eqnarray*}
Var(F|X_1^n)&=&\frac{4}{n^2}\sum_{i,j}(m_i-E\hat{m}_i)(m_j-E\hat{m}_j)Cov(\Delta_i^v\sqrt{v_i}\epsilon_i,\Delta_j^v\sqrt{v_j}\epsilon_j|X_1^n)\\
&=&o(\frac{4M_v}{n^2}\sum_{i,j}|Cov(\Delta_i^v\epsilon_i,\Delta_j^v
\epsilon_j|X_1^n)|).
\end{eqnarray*}
Thus $Var(F)=E(Var(F|X_1^n))=o(s^v_{n,2}/n^2)$ and $F=o(\sqrt{s^v_{n,2}\log n/n^2})$.

Also, for the second term in (\ref{Bdecomp}), we have $E(G)=0$ and
\begin{eqnarray*}
Var(G|X_1^n)&=&\frac{4}{n^2}Var(\sum_i\Delta_i^v\sqrt{v_i}\epsilon_i(\sum_jw_{ij}m_j-E\sum_jw_{ij}m_j)|X_1^n)\\
&\le&\frac{4M_m^2M_v}{n^2}\sum_{i,j}|Cov(\Delta_i^v\epsilon_i,\Delta_j^v\epsilon_j|X_1^n)|
\end{eqnarray*}
Thus $Var(G)=E(Var(G|X_1^n))=O(s^v_{n,2}/n^2)$ and $G=O(\sqrt{s^v_{n,2}\log n/n^2})$.
$\Box$
%\begin{eqnarray*}
%E(G)&=&\frac{2}{n}\sum_i\Delta_i^v\sqrt{v_i}\epsilon_i\sum_jw_{ij}\epsilon_j\\
%&=&O(s^v_{n,5}/n)
%\end{eqnarray*}
%and
%\begin{eqnarray*}
%Var(G|X)&=&\frac{4}{n^2}Var(\sum_i\Delta_i^v\sqrt{v_i}\epsilon_i\sum_jw_{ij}\sqrt{v_j}\epsilon_j|X)\\
%&=&\frac{4M_v}{n^2}\sum_{i,j,k,l}|Cov(\Delta_iw_{ij}\epsilon_i\epsilon_j,\Delta_kw_{kl}\epsilon_k\epsilon_l|X)\\
%%&=&O(s^v_{n,6}/n^2)
%\end{eqnarray*}
%Thus $Var(G)=O(s^v_{n,6}/n^2+s^v_{n,5}/n^2)$ and $G=O(s^v_{n,5}/n+\sqrt{s^v_{n,6}\log n}/n)$.

Finally, for the third term $H$, 
\begin{eqnarray*}
E(H)&=&\frac{2}{n}E\sum_i\Delta_i^v\sqrt{v_i}\epsilon_i\sum_jw_{ij}\sqrt{v_j}\epsilon_j\\
&=&O(s^m_{n,4}/n^2)
\end{eqnarray*}
and
\begin{eqnarray*}
Var(H|X_1^n)&=&\frac{4}{n^2}Var(\sum_i\Delta_i^v\sqrt{v_i}\epsilon_i\sum_jw_{ij}\sqrt{v_j}\epsilon_j|X_1^n)\\
&=&\frac{4M_v^2}{n^2}\sum_{i,j,k,l}|Cov(\Delta_i^vw_{ij}\epsilon_i\epsilon_j,\Delta_k^vw_{kl}\epsilon_k\epsilon_l|X_1^n).\\
%&=&O(s^v_{n,6}/n^2)
\end{eqnarray*}
Thus $H=O(s^m_{n,4}/n^2+\sqrt{s^v_{n,4}\log n}/n)$.

\textbf{Proof of Corollary \ref{cor:1}:}
We need to show that in the i.i.d. case, $s^m_n=O(n/\phi_m(h_m))$ and $s_n^v=O(n/\phi_v(h_v))$. 
We choose to calculate $s^m_{n,1}$, $s^m_{n,4}$ and $s^v_{n,4}$, the calculations are similar for the others.

In the i.i.d. case, we have
\begin{eqnarray*}
\sum_{i,j}|Cov(\Delta_i^m\epsilon_i,\Delta_j^m\epsilon_j|X_1^n)|&=&\sum_i{(\Delta_i^m)}^2.
\end{eqnarray*}
Thus $s^m_{n,1}=nE\Delta_1^2=O(n/\phi_m(h_m))$ by Lemma 4.3 of \cite{ferraty06}.

For $s^m_{n,4}$, we have
\begin{eqnarray*}
E\sum_{i,j}\Delta_i^vw_{ij}\epsilon_i\epsilon_j&=&E\sum_i\Delta_i^vw_{ii}\epsilon_i^2\\
&=&O\left(\frac{1}{n\phi_m(h_m)}E\sum_i\Delta_i^v\right)=O(\frac{1}{\phi_m(h_m)}),\\
\end{eqnarray*}
where we used the fact $w_{ii}=K(0)/\sum_jK(h_m^{-1}d_m(X_i,X_j))=O((n\phi_m(h_m))^{-1})$ obtained from (\ref{mean1}) and Lemma 4.3 of \cite{ferraty06}.

For $s^v_{n,4}$, we have 
\begin{eqnarray*}
&&\sum_{i,j,k,l}|Cov(\Delta_i^vw_{ij}\epsilon_i\epsilon_j,\Delta_k^vw_{kl}\epsilon_k\epsilon_l|X_1^n)\\
&=&\sum_i{(\Delta_i^v)}^2w_{ii}^2+2\sum_{i\neq j}{(\Delta_i^v)}^2w_{ij}^2\\
&=&O(\frac{1}{n\phi_m(h_m)\phi_v(h_v)})+O(\frac{1}{\phi_m(h_m)\phi_v(h_v)})\\
&=&O(n/\phi_v(h_v)),
\end{eqnarray*}
since it is assumed that $n\phi_m(h_m)\rightarrow\infty$.

\bibliographystyle{biometrika}
\bibliography{papers,books}

\end{document}